# Polarization fluctuation dominated electrical transport processes of polymer based ferroelectric-field-effect transistors

Satyaprasad P Senanayak[1], S. Guha[2] and K. S. Narayan[1]*

[1]Chemistry and Physics of Materials Unit

Jawaharlal Nehru Centre for Advanced Scientific Research

 Bangalore -560064 (India)

[2] Department of Physics and Astronomy, University of Missouri, Columbia, MO 65211, USA.

Corresponding Author: narayan@jncasr.ac.in

## Abstract

Ferroelectric field-effect transistors (FE-FETs) consisting of tunable dielectric layers are utilized to investigate interfacial transport processes. Large changes in the dielectric constant as a function of temperature are observed in FE-FETs in conjunction with the ferroelectric to paraelectric transition. The devices offer a test bed to evaluate specific effects of polarization on the electrical processes. FE-FETs have dominant contributions from polarization-fluctuation rather than static dipolar disorder prevalent in high *k* paraelectric dielectric-based FETs. Additionally, photo-excitation measurements in the depletion mode reveal clear features in the FET response at different temperatures, indicative of different transport regimes.

# I.  Introduction

Polymer field-effect transistors (FETs) are being actively developed for applications in the field of large area electronics.[1-2] Considerable efforts have been made to enhance the performance of these devices by combination of molecular and macroscopic processing approaches to obtain ordered microstructures and minimize the influence of the disorder prevailing in these systems.[3-4] Sustained efforts in this field have led to considerable improvement in polymer systems with high degree of self-organization and mobility ($\mu_{FET}$) magnitudes approaching 1 cm$^2$V$^{-1}$s$^{-1}$.[3, 5-6] The general mechanism of transport in these devices primarily arise from localized states in the framework of hopping transport and polaronic models.[7-9]

The dielectric layer plays an important role on the interfacial charge transport in polymer/molecular-based FETs.[7, 10-11] Effects of the relatively high dielectric constant (*k*) on the transport mechanism of molecular crystals have been studied from the perspective of Fröhlich polarons ,i.e., charge carriers bound to an ionic polarization cloud in the surrounding medium.[8] Unlike the studies on small molecule based FETs, there have been fewer efforts in polymer FETs to identify and evaluate the effects of polarization arising from the dielectric layer.[12-13] In case of macromolecular semiconductors, the model adopted to explain the influence of polar dielectric is in terms of a static dipolar disorder.[7-8] Based on these studies, it is largely accepted that low polarity dielectric interfaces are beneficial in minimizing localization enhancement and provide lower hysteresis, whereas highly polarizable dielectric induces higher charge density (*n*) and can consequently reduce the operating voltage.[14] Lower $\mu_{FET}$ for high *k* dielectrics is primarily attributed to the energetic disorder arising from static dipoles at the interface. The magnitude of broadening of the density of states ($E_{broad}$) varies in the range of 57 meV to 85 meV for *k* values between 2 - 4.[7] This dielectric induced energetic disorder has been studied by the temperature dependent transport in the framework of thermally assisted hopping.[7] Arrhenius type models have also been used to obtain activation energies ($E_A$) for polaron (Holstein) relaxation mechanisms and multiple trap levels.[7, 9] Additionally, there have been reports with electrolytes or ionic liquids as dielectric where the ultrathin Debye layer acts as a capacitive layer enabling device operation at low threshold voltages ($V_{th}$). The carrier density 'n' in these electrolyte gated FETs increase by several orders (> $10^{14}$ cm$^{-2}$) and hence the disorder induced trap states are filled to a greater extent even at smaller gate voltages ($V_g$).[15] $\mu_{FET}$ in these devices increase with $n$ since the carriers encounter fewer empty trap sites and the transport approaches metallic conductivity with low activation energy ($E_A$ ~ 7 meV compared to $E_A$ ~ 56 meV observed in case of low $k$ dielectric for regioregular-poly-3-hexyl thiophene - P3HT).[10]

We explore the influence of polar ferroelectric (FE) layer on the polymer transport layer by carrying out a systematic set of studies at different temperatures (*T*). A FE dielectric layer, which permits the access to nearly an order of magnitude range of polarization (k ~ 4 to 24) with *T* as the tuning parameter offers a test bed for following the changes in transport physics. Such a large variation in *k* permits tuning of the polarization strength from weak to strong coupling regime; in the latter case, there is stronger interaction between the interfacial carriers and the dielectric polarization. We also focus on the high *T* regime where a crossover from the ferroelectric to paraelectric (PE) regime is present. The choice of a FE layer with an accessible transition temperature ($T_c$) beyond which the layer becomes a regular PE layer provides a suitable platform to study the effect of surface polarization on the transport channel in a controlled manner. Changes arising due to the phase transition from a polar FE state to a non-polar PE state appear as distinct signatures in the FET characteristics. A single dielectric layer with tunable surface polarization by varying *T* allows a clean diagnostic tool to understand the transport compared to using different dielectrics with varying dielectric constant, where processing condition, morphology and interfacial energy plays a significant role.

The dielectric materials of choice are Polyvinylidene Fluoride (PVDF) homopolymer and PVDF based copolymers like P(VDF-Hexafloropropylene) (PVDF-HFP) and P(VDF-Trifloroethylene) (PVDF-TrFE). PVDF based dielectrics are known to exist in three regular conformations corresponding to its phases with similar potential energies.[16] The three known conformations correspond to *all-trans(β-phase)*, $tg^+tg^-$ *(α-phase)* and $tttg^+tttg^-$ *(γ-phase)* with *β-phase* being the most polar FE-phase. The phase transition in this dielectric is an order-disorder structural transition where the *β-phase* is converted to *α-phase*. Incidentally, the *β-phase* of PVDF-TrFE has been recently used to enhance the efficiency of organic solar cells[17] where the long-range FE-field enhances the charge carrier separation. These FE-dielectric materials develop a depolarizing field ($\varepsilon_{dep}$) in the presence of a semiconductor.[18-19] The $\varepsilon_{dep}$ at the semiconductor-FE interface is overcome to a large extent by photo-generated carriers.[20] Tunability of $\varepsilon_{dep}$ in the present device geometry can be used to observe the difference in dynamics of photo-generation in polar and non-polar state of the dielectric.

The FE environment for PVDF-based polymer dielectrics at the interface also acts as a source of fluctuation, which is reflected in strong *T* dependence of *k* in the ordered phase over a wide frequency range. The thermalization of individual dipoles result in a negative pyroelectric coefficient or a smaller remnant polarization ($P_r$) at the interface as the system approaches $T_c$ which needs to be taken into account to accurately estimate $\mu_{FET}(T)$. Additionally, thermal fluctuations of the ferroelectric domains translate to a characteristic macroscopic dipolar-disorder at the charge transport interface, which is evident from our experimental results. The FE-FET system thus provides a platform to achieve a tunable polarization enabling a better understanding of the interfacial charge transport with minimum errors.

## II. Experimental Section

**Materials**: Conjugated polymer rr-P3HT (weight average molecular weight, $M_w$ ~87,000) was procured from Sigma Aldrich Inc., Dielectric materials PVDF ($M_w$ ~ 180,000); PVDF-HFP copolymer ($M_w$ ~ 455,000), PVDF-TrFE (75/25) and Hydroxyl free divinyltetramethylsiloxane bisbenzocyclobutene (BCB) were obtained from Sigma Aldrich Inc., Measurement Specialties Inc., USA and Dow Chemical respectively.

**Device Fabrication**: Bottom gated top contact FETs were fabricated by coating required electrode-Al or Cr/Au ($10^{-6}$ mbar, 1 $A^0$/s, 30 nm thick) by shadow mask technique on standard RCA cleaned glass substrates. Then a dielectric layer was obtained by spin coating the polymer dielectrics in their respective solvents (PVDF and PVDF-HFP at 80mg/ml in N, N-Dimethylacetamide, PVDF-TrFE (75/25) at 40mg/ml in Butane-2-one and BCB in Mesitylene) at 1000 rpm for 1 minute to obtain films of thickness 0.6-0.8 μm. The films were rapidly annealed in vacuum below their melting point (150°C for PVDF, 120°C for PVDF-HFP and PVDF-TrFE) to improve crystallinity of the films and obtain desired phases of the polymer. It has been shown that PVDF thin films prefer *β-phase* on a gold coated substrate when annealed at $150^0$C for 2 hrs.[21] In order to enrich *β-phase*, FETs were fabricated with gold as the gate electrode. In the cases where Al was employed, mixed phases obtained are typically not ferroelectric. For copolymers like PVDF-TrFE, due to the greater proportion of the bulkier fluorine atoms, the molecules cannot be accommodated in *tg$^+$tg$^-$* conformation and hence they crystallize in all-trans form.[16] The BCB films were annealed for 30 minutes at 290 $^0$C in vacuum. P3HT was then coated from a solution of 10mg/ml in chlorobenzene at 1000 rpm for 1 minute to obtain films of thickness ~ 80nm. This was followed by the deposition of Au source drain electrode ($10^{-6}$ mbar, 1 $A^0$/s, 20 nm thick) by shadow masking technique to obtain channels of length 80-100 μm and width 1mm (shown in **Figure 1a**).

Bi-layer dielectric based FETs were fabricated with BCB layer in contact with the gate-electrode and PVDF based dielectric layers in desired phases interfacing the semiconductor layer. Thicknesses of the layers of dielectric materials were appropriately chosen to have larger contribution to the capacitive coupling from the BCB component.

**Electrical characterization**:

**FET**: Standard bottom gated FETs were characterized with two identical source meters - Keithley 2400 and high impedance electrometer Keithley 6514 and cross-checked with measurements from a standard Keithley 4200 semiconductor parameter analyzer. The devices exhibited *p*-type transport with current modulation > $10^3$ -$10^4$ (leakage current three to four orders of magnitude lower). Issues related to roughness and porosity of PVDF based dielectric films were resolved systematically,[22] and the dielectric quality was comparable to recent reports on bottom gated top contact FETs made with these dielectrics, which have demonstrated reliable and leakage-free transport, $\mu_{FET}$ ~ 0.01 cm$^2$V$^{-1}$s$^{-1}$ in similar geometry.[23] More than 100 devices were tested for these studies, and the results presented are from representative sample devices, which constitute the statistical median.

**Figure 1c** depicts the typical transconductance curve for the top contact bottom gated transistors made with P3HT as the active semiconducting layer and PVDF-TrFE as the dielectric. In order to ascertain that the current variation is not a consequence of factors such as bias stress and process variations, the devices were made to go through a complete thermal cycle (100 K – 450 K). Typical applications with FE-FET involve hysteresis and memory as important phenomenon. In this study hysteresis effects were minimized to a large extent by applying sufficient positive bias and driving back the FET to the original off state condition (within 0 - 5 % variation) to have independent measurements at each *T*. Similar *T* dependant studies and transconductance measurements were also done on bi-layer dielectric based FETs.

**Capacitance measurement**: The capacitors fabricated alongside these FETs as M-I-M (Metal-Insulator-Metal) structures were characterized at 100 Hz using HP4294A (**Figure 2**). FE-phase of PVDF-TrFE and $\beta$-PVDF were verified experimentally by *P-E* (Polarization-Electric field) profiles, while PVDF-HFP did not indicate FE traits. $T_c$ ($\approx$ 390 K) of PVDF-TrFE was verified by $C(\omega,T)$ measurements and X-ray measurements, and is consistent with reported values.[24] $\beta$-PVDF continues to be in the FE phase in the measurement range ~ 360 K. The $C(T)$ measurements of the bi-layer indicated only a marginal variation of $k(T)$ (or minimal polarization fluctuation with *T*) similar to that of a single layer BCB capacitor. The temperature dependence studies were carried out using liquid helium based cryostat from Cryogenics Technology Ltd.

**Raman spectroscopy**: Raman Spectra was collected by an Invia Renishaw spectrometer attached to a confocal microscope with 50× long objective using 785 nm line of a diode laser as the excitation wavelength. The samples were affixed to a stainless steel sample holder of a Linkam LTS350 microscope hot-cold stage.

**Photo-excitation measurement**: For the photocurrent ($I_{ph}$) measurements FE-FETs were made to operate in the depletion mode at $V_g$ = 20 V, drain source bias ($V_d$) = -60 V and CW laser of 532 nm with power 2mW/cm$^2$ was used to illuminate the channel completely at T ~ 180 K and 400 K. $I_{ph}(t)$ was recorded using high impedance electrometer Keithley 6514.

### III. Results and Discussions

#### A. Non-linear polarization contribution in FE- FET

In a FE dielectric, the total polarization ($P_{tot}$) has a non-linear contribution in addition to the linear contribution: $P_{tot} = P_{lin} + 2P_r - P_{sat}$,[25] where $P_{lin}$ is the linear contribution to the polarization, $P_r$ is the remnant polarization and $P_{sat}$ is the saturation polarization. At low *T*, $P_{tot} \approx P_{lin} + P_r$ (as $P_r \sim P_{sat}$) and for high *T*, $P_{tot} \approx P_{lin} - P_{sat}$ (as $P_r \ll P_{sat}$). The measured $k(T)$ (**Figure 2**) of the dielectric films used for the present study reflects thermally activated relaxation mechanism of the polymer chain which promotes the ordered FE-phase. This enhances the dielectric constant and hence the *k* variation in the 200 K – 420 K regime. *k(T)* is a measure of the order parameter in the FE phase and takes a quadratic form. The bulk *k* is representative of the surface polarization since uniform phases exists both at the bulk and surface for such thick films. *k(T)* is a robust function and is not sensitive to poling treatment normally applied to the films to yield higher *k* values in this voltage range of polarization saturation of the dielectric. It is to be noted that conventional FET equations do not account for the additional non-linear polarization ($P_{nl} = 2P_r + P_{sat}$) of the dielectric. A compact model of a FE-FET involves an equivalent circuit at the gate consisting of one resistor and four capacitors in parallel (one of which is linear and other three being non-linear).[26] In the framework of distributed threshold switching model for PVDF the relaxation period of domains are distributed.[26] We take into account the above factors and approximate the treatment in the present case by including a correction to $V_g$, in the form of an equivalent voltage due to FE-polarization: $V_0 \approx P_{nl} d / \varepsilon_0 \varepsilon_r$ where *d* is the thickness of dielectric layer, $\varepsilon_0$ is the permittivity of vacuum, and $\varepsilon_r$ is the dielectric constant of the dielectric layer. This voltage accounts for the additional charge density at the channel. The saturation regime $\mu_{FET}$ was calculated with the modified transconductance expression of the drain source current in the saturation regime: $I_{ds} = (\mu_{FET} W C_0 / 2L)(V_g + V_0 - V_{th})^2$, where $C_0$ is the effective capacitance of the dielectric layer, *W* is the channel width and *L* is the channel length of the transistor (see Appendix 1).

The experimentally measured polarization and *k* at different *T* is incorporated into the modified FET equation to arrive at $\mu_{FET}(T)$ as shown in **Figure 3**. $V_0$ obtained from the drain current, $I_d(V_g)$, at different *T*, indeed follows a similar trend as $P_r(T)$ (independently obtained from *P-E* measurements). $P_{nl}$ translates to a small correction in the final $\mu_{FET}$ and minor discrepancies in the estimate may arise if poling is not carried out during the film forming process. Corrections arising from the $P_{nl}$ differs from the effective gate voltage correction due to factors originating from semiconductor carrier density[27] and trapping of the induced charges at low $T$. It was ensured that the value of $C$ obtained from $C$-$V$ measurements was in a similar bias range as that employed for extracting $\mu_{FET}$. We note that $k(T)$ represents a macroscopically averaged quantity for estimating $\mu_{FET}$ which is essentially controlled by microscopic parameters. The inherent variations and fluctuations of the bulk dielectric can dominate $\mu_{FET}(T)$ response and may not reflect the real transport behavior. Hence fluctuations can be represented by $<\mu_{FET}(T)>$.

## B. Temperature dependant mobility estimation

$<\mu_{FET}(T)>$ is obtained for different systems in the range 100 K $< T_c <$ 420 K (**Figure 3**). Arrhenius fits are used to evaluate the $T$-dependence. Different dielectrics yield the following transport characteristics from $T$-dependent FET studies: (i) weak $T$ dependence of $<\mu_{FET}(T)>$ in the FE regime, ($E_A <$ 15 meV) for PVDF-TrFE based FE-FETs, which becomes strongly $T$ dependent upon the phase transition to PE regime ($E_A \sim$ 0.4 eV); (ii) for $\beta$-PVDF based FETs, $<\mu_{FET}(T)>$ is nearly $T$ independent, and in fact appears to marginally decrease with increasing $T$ ( 240 K $<$ T $<$ 360 K) (Supplementary Section); (iii) devices with high $k$ paraelectric non-FE dielectric (PE-FET) exhibits a strong $T$-dependence $<\mu_{FET}(T)>$ ($E_A \sim$ 46-76 meV) in 200 K $<$ T $<$ 290 K regime.

The apparent weak $T$ dependence of $<\mu_{FET}(T)>$ cannot be construed as an indicator of higher delocalization lengths since the fluctuations due to $k(T)$ are inbuilt in the $<\mu_{FET}(T)>$ estimates. Hence, $E_A$ obtained from $\mu_{FET}(T)$ should not be taken as a figure of merit to gauge the order (band-like versus hopping) in such device systems. It has been reported that in PVDF the FE switching usually involves domains, which is then expected to be accompanied by $T$ dependent relaxation processes and can manifest as fluctuations. In case of FE-FETs, the observed $<\mu_{FET}(T)>$ includes contribution from $I_{ds}(T)$, thermal motion of the dipoles gauged by the pyroelectric coefficient, and collective domain fluctuations. In general, $I_{ds}(T)$ is an increasing function of $T$, thermal vibration of dipoles decreases $P_r$ with $T$ (**Figure S1**), and the domain fluctuation increases with $T$ ($T < T_c$). It can be inferred from the $<\mu_{FET}(T)>$ behavior that the transport is dominated by macroscopic fluctuations (with minor contribution from the thermal motion of individual dipoles) within the FE-regime creating a macroscopic dipolar disorder at the interface and suppressing the observed $E_A$.

The dynamic disorder arising from the fluctuations appears to be more significant in the case of $\beta$-PVDF which is less crystalline than PVDF-TrFE.[16] Energetically, the variations in polarization can result in relative broadening of the density of occupied states (DOOS) creating a disordered landscape for charge transport. In case of the high-$k$ (PE) dielectrics with random dipoles, this form of disorder appears to play a smaller role than in the case of FE dielectrics. In PE-dielectrics, the inherent static dipolar disorder dominates the collective fluctuation due to $T$. Upon comparing different forms of PE-PVDF based FETs, the random fluctuation as indicated by the $E_A$ value is significant in the case of largely amorphous $\alpha$-PVDF. It should be noted that this effect of polarization fluctuation seen as variation in $k(T)$ is different for FE and PE dielectric regimes. The effect of $k(T)$ variation on $I_{ds}(T)$ is significantly more in FE-FET than the regular high $k$ PE dielectric based FETs and becomes a dominant factor in the FE regime. The distinction between collective (FE) and non-collective polarization (regular high-$k$) processes is equivalent to the difference between coherent motion of a large number of molecules and non correlated displacements.[28] The transition becomes obvious upon comparing the $<\mu_{FET}(T)>$ in the region above and below $T_c$ as in the case of PVDF-TrFE. It can thus be concluded that the macroscopic polarization fluctuation is more dominant in FE-FET and even higher, when the dielectric is relatively more amorphous.

## C. Temperature dependant mobility estimation for FETs with bi-layer (low-k|high-k) dielectric

In order to distinguish the thermal processes and polarization contribution to $<\mu_{FET}>$, bi-layer dielectric based FETs were fabricated. Bi-layer dielectric based devices provide a clear distinction between the P3HT-FE and P3HT-PE interfaces. The features observed from $T$ dependent studies of FETs with different bi-layer of dielectrics are the following: (i) $\mu_{FET}^{BCB/FE} > \mu_{FET}^{BCB} > \mu_{FET}^{FE}$ at 300 K where $\mu_{FET}^{BCB/FE}$ is the mobility for FETs with BCB|PVDF-TrFE as the dielectric, $\mu_{FET}^{BCB}$ and $\mu_{FET}^{FE}$ are the field effect mobilities for BCB-based and FE-PVDF-TrFE-based FETs respectively; (ii) for devices with BCB|PVDF-TrFE in the FE regime, the $\mu_{FET}(T)$ has an activated behavior with activation energy, $E_A^{bi-layer} \sim 75$ meV, which has a stronger $T$ dependence upon phase transition to the PE-regime ($E_A^{bi-layer} \sim 130$ meV); (iii) bi-layer dielectric devices made with BCB|high $k$ PE-dielectrics also showed an activation mechanism ($E_A^{bi-layer} \sim 140$ meV) higher than the single layer dielectric devices. It was not possible to make bi-layer dielectric devices with BCB|β-PVDF because dominant β-phase requires the growth of PVDF films on gold substrates.

All the devices made from the bi-layer dielectrics exhibited sizable activated behavior and is not dominated by the FE characteristic of fluctuation with $T$. This can be explained on the basis of a reduced polarization of the FE layer. The $k(T)$ and switching characteristics (in FE dielectric) have been analyzed in terms of nucleation and growth model.[28] The process of kink propagation required for the domain growth provides the source for fluctuation. It was established that this nucleation and growth process decreases with decreased bias and this trend possibly explains the present case of minimal effect from the FE layer in the bi-layer geometry. The experimental results points to a conclusion that bi-layer dielectric device can be utilized to overcome the polarization fluctuation, which is prevalent in case of the single layer FE-FET. This outcome can be of significance in making devices with sustainable parameters over a wide range of temperature.

Another observation from the $T$ dependent study is the lowering of $E_A^{bi\text{-}layer}$ with FE interface. This is indicative of inherent statically ordered phase at the interface similar to the case of low $k$ BCB dielectric. The impact of the static dipolar disorder observed in high $k$ PE-dielectric is apparently mitigated by employing a FE-dielectric. The transport mechanism is then the combined result of the associated disorder originating from these contrasting environments. The decreased $E_A^{bi\text{-}layer}$ with FE interface can also be attributed to higher carrier concentration, which is consistent with a continuum of localized states above a (hole) transport level, $E_t$. As the states above $E_t$ are filled with increasing $n$, the quasi-Fermi level ($E_F$) is lowered and $E_A$, i.e., $E_F$–$E_t$ decreases.[11]

### D. Analysis of temperature dependence

If FE component of the bi-layer device is assumed to contribute negligibly to the transport then $<\mu_{FET}^{bi\text{-}layer}(T)>$ can be used as the background reference to evaluate contribution from other additional factors. The polarization fluctuation dependent mobility $\mu_P$ can be obtained by expressing $<\mu_{FET}>$ as a combination (Matthiessen's rule) along with corrections arising from different parameters related to interfacial properties and charge retention at the channel by the FE-dielectric, i.e.,

$$<\mu_{FET}> = \left(\frac{1}{<\mu_{FET}^{bi-layer}>} + \frac{1}{<\mu_P>}\right)^{-1} + \frac{k'(V_g-V_t)^{\delta+1}}{\gamma+1}$$

where, $<\mu_{FET}^{bi\text{-}layer}>$ is the $T$ dependent mobility of bi-layer dielectric based FETs, $k'$ is related to the ease of hopping which gives the interfacial properties (interfacial roughness, de-mixing and polymer film morphology on the dielectric), $\delta$ is related to the characteristic temperature $T_0$, $\gamma$ is related to the width of DOS.[29] This model can be utilized to independently extract polarization dependent $<\mu_{FET}>$. This polarization dependent $<\mu_{FET}>$ can deviate from an Arrhenius type behavior[23].

### E. Temperature dependent study of FE-FET parameters ($V_{th}$, $\sigma$, $S$)

The mobility measurements in FE-FETs are supplemented by conductivity measurements ($\sigma(T)$) to get a deeper insight into the nature of interfacial transport. $\sigma(T)$ is obtained in the linear regime of operation of the FE-FET with PVDF-TrFE as the dielectric. In the context of 2-D variable range hopping model, $\sigma \propto \exp(T_0/T)^{1/2}$; the localization radius obtained from the Efros-Shklovskii type transport[30] was in the range of 1.5 – 1.9 nm, a value in between that of a PE dielectric (0.7 nm) and an electrolyte (2.6 nm). $\sigma \sim 0.01$ S/cm points to a fraction ($\sim 0.1\%$) of conductivity contributed from delocalized charge carriers.[31]

The phase transition of the dielectric in single layered FE-FETs is directly reflected in $V_{th}$ (**Figure 4**). The sizable step increase in $V_{th}$ is also accompanied by a discontinuity in the sub threshold swing–$S$ (**Figure 4**). At $T_c$, the discontinuity in conductivity ($\approx \Delta\sigma$) is lower than $\Delta\mu_{FET}$ which implies a lowering of $n$, i.e., presence of remnant carriers enhance the channel conductance in the FE phase. $\Delta V_{th} \sim 12$ V at $T_c$ can then be reconciled and related to the magnitude of $\Delta n$. This clearly indicates the contribution from the polarization in the FE phase to $n$, which steeply decreases beyond $T_c$ (**Figure S3**).

$V_{th}$ studies as a function of $k$ and $T$ can differentiate the contribution of trapping from the impurities and polarization factors. An increase in $T$ is accompanied by a shift in $V_{th}$, attributed to increased $<\mu_{FET}>$ and decreased trapping,[32-33] while an increase in $k$ results in a higher density of dipolar disorder induced trap states and opposes the $V_{th}$ shift. In the present case, variation of $V_{th}$ with $T$ follows the $<\mu_{FET}>$ behavior in the entire FE-regime as is apparent from **Figure 5 (a)**. Since $V_{th}(T)$ does not exhibit any obvious relation with $k(T)$ the real picture is more complex than a simplistic impurity/disorder induced trapping, and reinforces the interpretation of the dominant role of random thermal fluctuations.

### F. Raman measurements

The *T* dependent Raman measurements verified the stability of P3HT (200 < T < 420 K) and is consistent with reported observations.[34-35] There have been reports where the FE transition of PVDF has been followed by Raman measurements.[36] The spectra of P3HT coated on PVDF-TrFE retained the features of the constituents, similar to P3HT on Si as a function of *T*. There was a change in the background of the Raman spectra in the PE-phase, which was reversible as the temperature was cycled to the FE-phase. However, the broadening and frequency shifts of the P3HT Raman peaks as a function of *T* were identical for P3HT on Si and PVDF-TrFe, indicating that the changes in the transport properties at the $T_c$ arises mainly from the coupling of charge carriers with the dielectric interface. **(see Supplementary Section).**

**G. Dynamics of depolarizing field monitored by photocurrent transient measurements**

In order to closely probe the effect of phase transition on transistor operation, the FETs were studied in the depletion mode under a constant photo-excitation **(Figure 6)**. When a transistor operates in depletion mode, the semiconductor does not compensate charges, resulting in depolarization of the ferroelectric material. As soon as light is incident on the semiconductor, photo-generated charge carriers provide the necessary compensating charge carriers enabling polarization. The photocurrent, $I_{ph}(t)$ response for FE-FET with PVDF-TrFE as the dielectric was distinctively different from the response obtained with P3HT FET with polyvinyl alcohol (PVA, high *k* ~ 10) as the dielectric.[37] The rate of increase in $I_{ph}(t)$ upon photo-excitation in FE-FETs is slower since a fraction of the photo-generated charge carriers are utilized to compensate $\varepsilon_{dep}$, and this contribution was clearly absent in the PE-phase above $T_c$. The drain source current with illumination can be described as: $I = I_o(V_g) + I_{ph}$, where $I_o$ is the dark value of the drain source current and $I_{ph}$ is the light induced component consisting of excess charge carrier generation rate and recombination processes. In the case of FE-FETs there is an additional contribution to the drift-term from the $\varepsilon_{dep}$, which has a comparatively slow relaxation.[18-20] Upon illumination at low $T$ ($T < 200$ K), a consistent trend of slower rising rate of $I_{ph}(t)$ with increasing $V_g$ (which implies increased depletion or higher $\varepsilon_{dep}$) is observed. At high $T$, ($T > 300$ K) the trend in the rising rate $I_{ph}(t)$ with continuous illumination is altered and carrier recombination dominates over depolarization factors which is $V_g$ independent.

The $I_{ph}$ decay response upon termination of a light pulse in general is a recombination limited process in the presence of large density of traps.[37] $I_d$ settles at an intermediate metastable state with an extremely slow relaxation (50 % drop over a period of 3 h). The retention factor, $R_f \approx I_d$ in the metastable state/$I_{dark}$, for devices with FE-dielectric is about 3 times more compared to devices where PVA was used. The high retention fraction is attributed to a slower relaxation of ferroelectric materials to the depolarized state in the absence of any external bias. $R_f$ decreases with increasing $T$ as expected and does not exist in the PE-phase since recombination factors dominate. Our observations from photo-generated transport show the existence of a field originating from the non-linear contribution to polarization - $P_{nl}$ of the FE-layer, which affects bulk charge transport in the polymer. This is a clear signature of a long-range polymer-dielectric interaction and is consistent with the physical model speculated for ferro and paraelectric dielectrics.

The change in the transport mechanism of FE-FET beyond $T_c$ is clearly signified in all the measurements involving the parameters $V_{th}$, S and σ as well as the photoinduced response. The underlying feature in the FE-FET is the weak activation at low T which gets magnified at T > $T_c$. The role of the FE layer becomes clearer from the studies of devices with bi-layer dielectrics. The dominant contribution from the bulk of dielectric emphasizes the role of polarization fluctuation on the transport in FE-FETs. This fluctuation plays a major role and is accompanied by a smaller contribution from $P_{nl}$ in the FE-regime.

## IV. <u>Conclusions</u>

In summary, $k(T)$ in the FE-regime and the phase transition at $T_c$ is reflected in the accompanying FET characteristics. The FET parameters $\mu_{FET}$ and $V_{th}$ offer a handle to probe transport in these devices. The transport in FE-P3HT FETs clearly depends on polarization to a large extent, unlike the FETs with non-polar dielectrics where the dipolar induced energetic disorder is dominant. The evidence for this interpretation was arrived from the weak T dependence of the different electrical parameters in the FE regime compared to the strongly activated behavior in the PE regime. Further, bi-layer dielectric layers, which mitigate the influence of the polarization of single layer dielectric based FETs were studied and compared. The absence of dominant contribution from the bulk-semiconductor polymer layer was verified by studying normal dielectric FETs. T dependence of Raman and X-ray features also confirmed the stability and integrity of the bulk semiconductor layer. In addition, photocarrier-generation studies in the depletion mode of the FET provide a clear distinction in the transport phenomenon in conjunction with the phase transition of the dielectric. The charge carriers in the FE-phase experience an additional interaction originating from the FE-polarization. The distinct features in the FE regime are then attributed to the considerable influence of dielectric polarization fluctuations, which dominates over conventionally observed intrinsic processes.

**Appendix 1**: **Transistor equations derivation for FE-FET:**

Basic Assumptions for the derivation -:
1. OTFTs with p-type semiconductor and P(VDF-TrFE) / β-PVDF gate dielectric.
2. The ferroelectric layer is not pre-poled by applying $V_S = V_D = 0V$; $V_G < 0V$
3. The FE-FET also showed remnant polarization in the transconductance measurements. The remnant polarization, $P_r$, is attributed to dipoles in crystalline phase of P(VDF-TrFE). The total polarization involves contribution from dipoles in the amorphous phase, which is proportional to the electric field inside the medium.

4. The poling process is assumed to be stable in the poling condition, which means further poling does not have additional effects on $P_r$; and $P_r$ is considered to be constant under variations of small electric fields.

The total polarization due to the ferroelectric dielectric has non-linear contribution in addition to the linear polarization

$P_{tot} = P_{lin} + P_{nl} = P_{lin} + 2P_r - P_{sat}$,

where, $P_{lin}$ is the linear contribution to polarization

$P_{nl}$ is the non-linear contribution to polarization

$P_r$ is the remnant polarization

$P_{sat}$ is saturation polarization.

In our case we have approximated

At low $T$, $P_{tot} = P_{lin} + P_r$ (as $P_r \sim P_{sat}$)

High $T$, $P_{tot} = P_{lin} - P_{sat}$ (as $P_r \ll P_{sat}$).

Charge at the semiconductor-dielectric interface is given by

$Q = \varepsilon_0 \varepsilon_r (V_i + V_o)/d$ (1)

and $V_0$ is deduced from $P_{nl} = \varepsilon_0 \varepsilon_r V_0/d$, where

$\varepsilon_o$ is the permittivity of the vacuum and $\varepsilon_r$ is the relative dielectric constant of the dielectric

Since $V_i$ can also be expressed as $(V_g - V_x)$:

$Q = C(V_g + V_o - V_x)$ (2)

$I_{ds}$ is thus calculated over the entire channel in the saturation regime with the assumption that semiconductor thickness $d \ll L$.

$$I_{ds} = \mu W/L \int_0^{V_d} Q dV_x$$

Substituting for $Q$ and with the approximation $V_d \sim V_g + V_o$

$$I_{ds} = \mu C \frac{W}{L} (V_g + V_0 - V_t)^2$$

## **Acknowledgments**

KSN acknowledges Alessandro Troissi for useful discussions, DAE Government of India, for funding and S.P.S acknowledges CSIR, India for the research fellowship.

**Supporting Information**

Supporting information contains variation of polarization retention with *T* obtained from P-E measurements, Raman spectra variation with *T*, equivalent circuit diagram for the depolarizing field, Output curves for FE-FET with PVDF-TrFE to obtain $\sigma(T)$. We also present $<\mu_{FET}>$ variation with *T* with ferroelectric β-PVDF dielectric and P3HT as the active semiconducting material. Device schematic, output and transconductance curves for bi-layer dielectrics FETs and hysteresis curves with different combination of dielectrics are also presented.

**Main Text Figures**

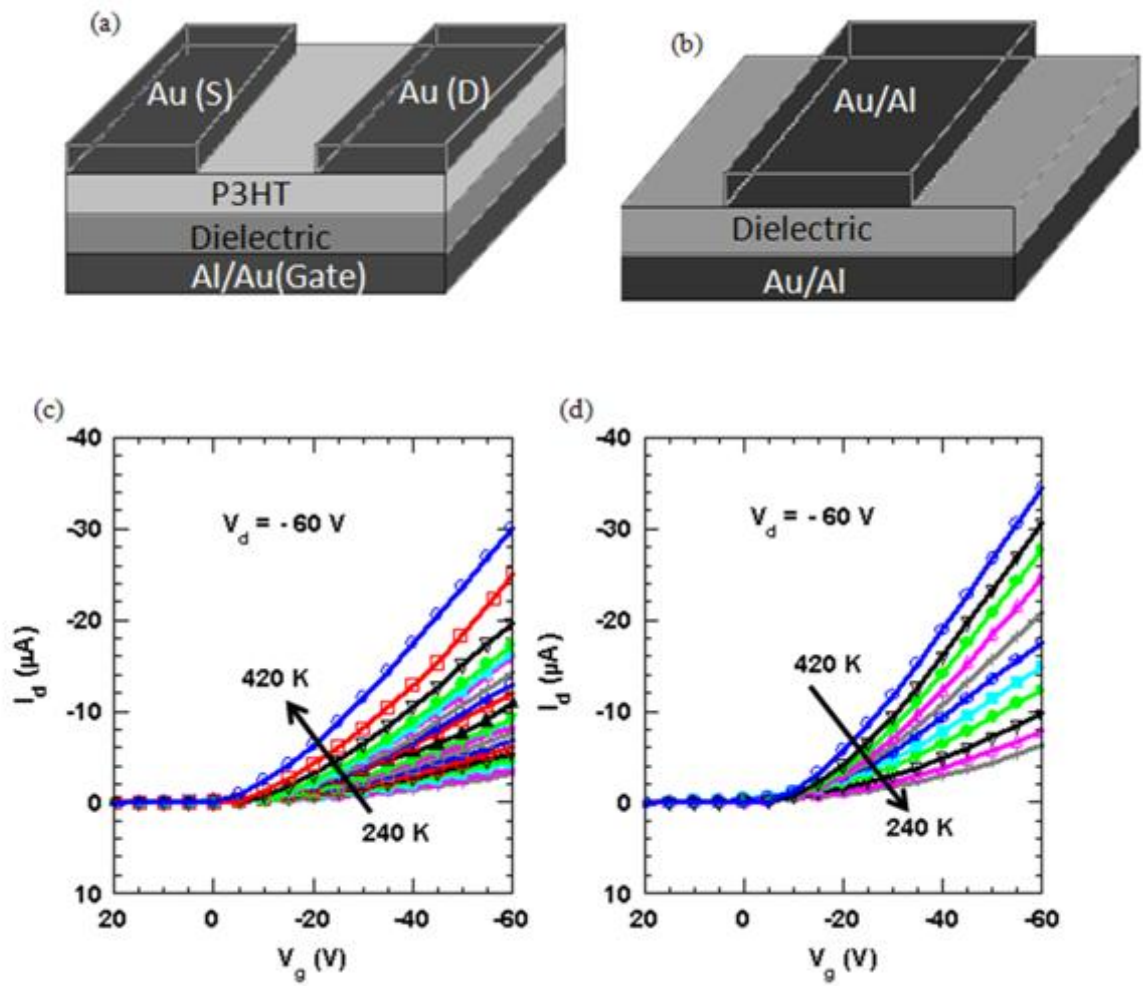

**Figure 1.** (a & b) Schematic of the top contact bottom gated FET and M-I-M structure for capacitance measurement. (c & d) Typical transconductance curves for PVDF-TrFE (75/25) based FE-FET for $T$ (240 K – 420 K) at $V_d$ = -60 V.

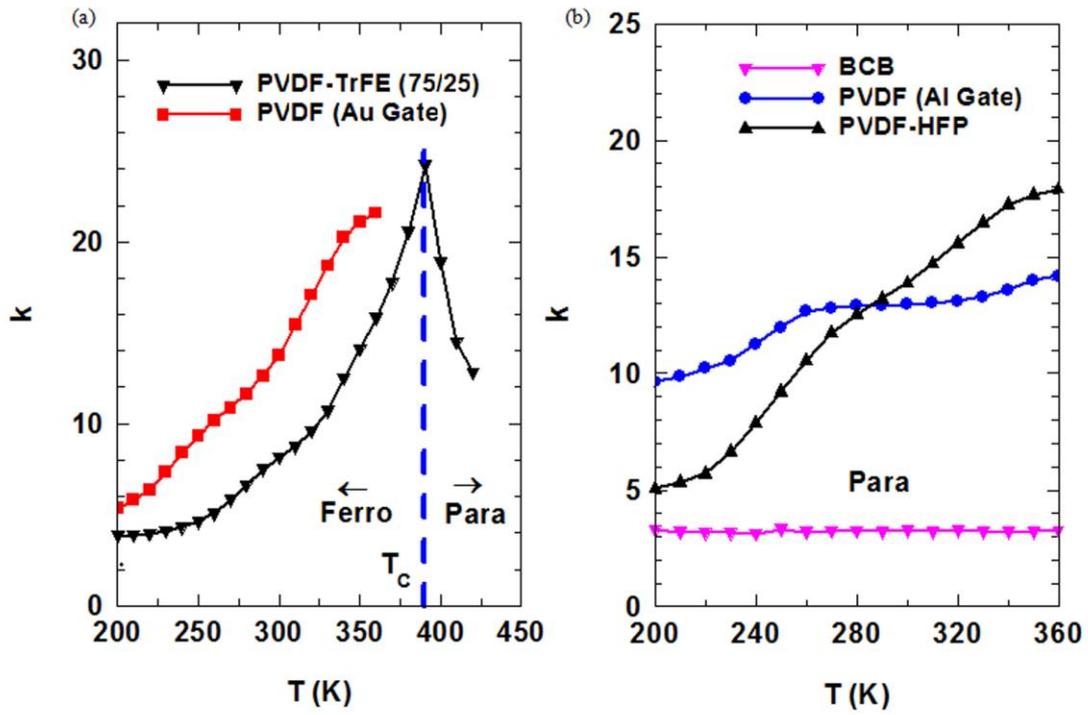

**Figure 2.** *k*(*T*) variation for different dielectrics measured at 100 Hz in M-I-M structure. PVDF-TrFE shows a clear phase transition from the ferroelectric to paraelectric regime.

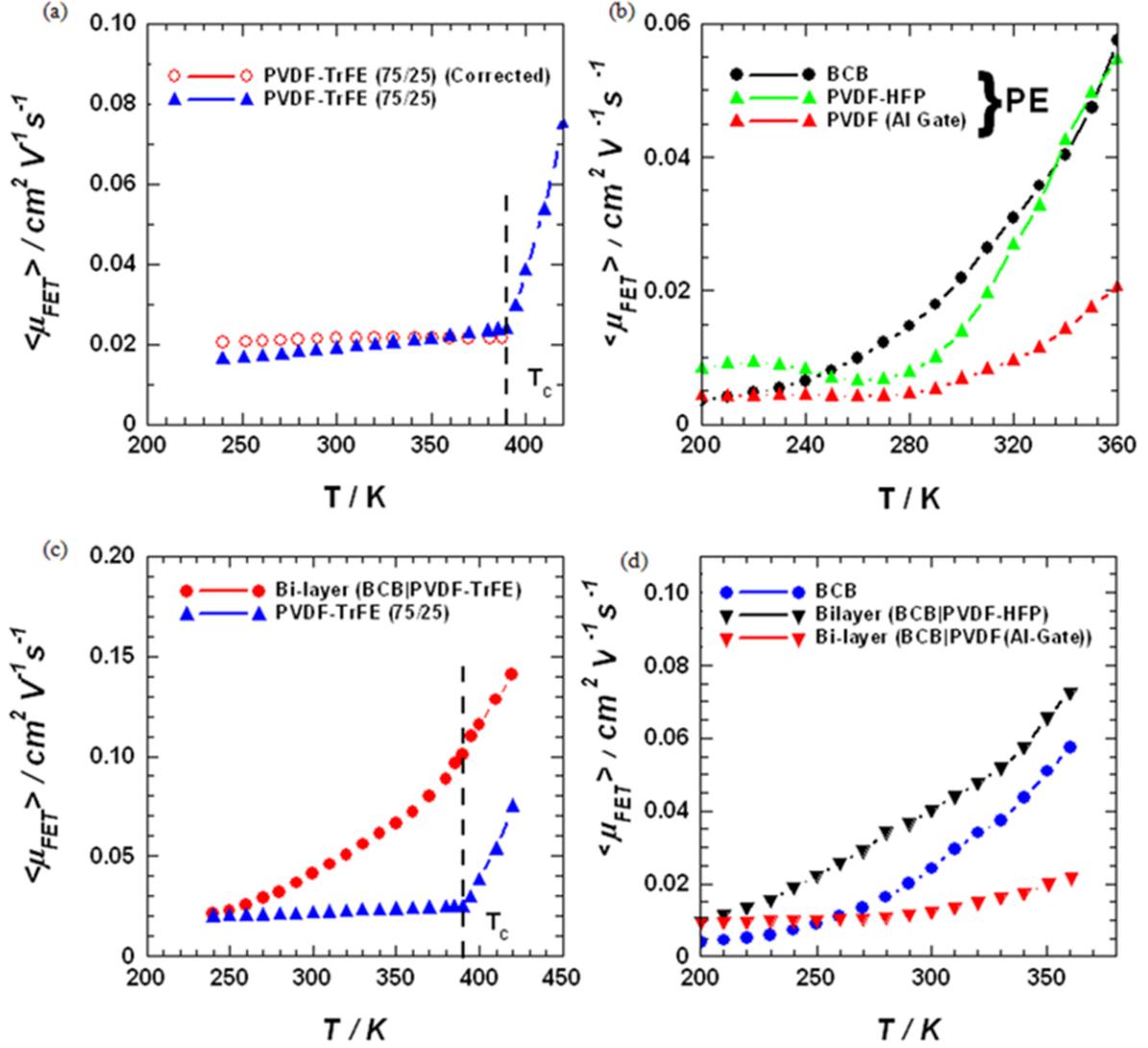

**Figure 3.** $<\mu_{FET}>$ for P3HT-FET with different dielectrics for devices having similar $W/L$ values. (a) Ferroelectric PVDF-TrFE showing a clear discontinuity in $<\mu_{FET}>$ at $T_c$ (~ 390 K). Solid lines indicate transport in the FE- regime and dashed line is for the transport in PE-regime. Corrected $<\mu_{FET}>$ is obtained by accounting for the nonlinear contribution to the ferroelectric polarization. (b) $<\mu_{FET}>$ versus $T$ for PE dielectrics PVDF-HFP, $\alpha$-PVDF and BCB, and (c,d) for bi-layer dielectrics with FE and PE interfaces.

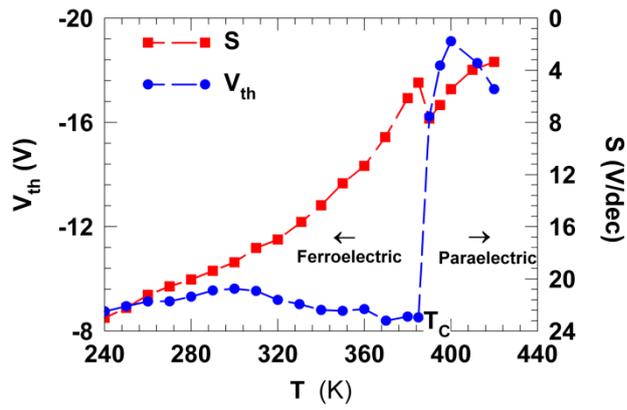

**Figure 4.** Variation of $V_{th}$ and $S$ with $T$, showing the effect of ferro to paraelectric transition obtained from a transistor with PVDF-TrFE as the dielectric and P3HT as the active semiconducting layer.

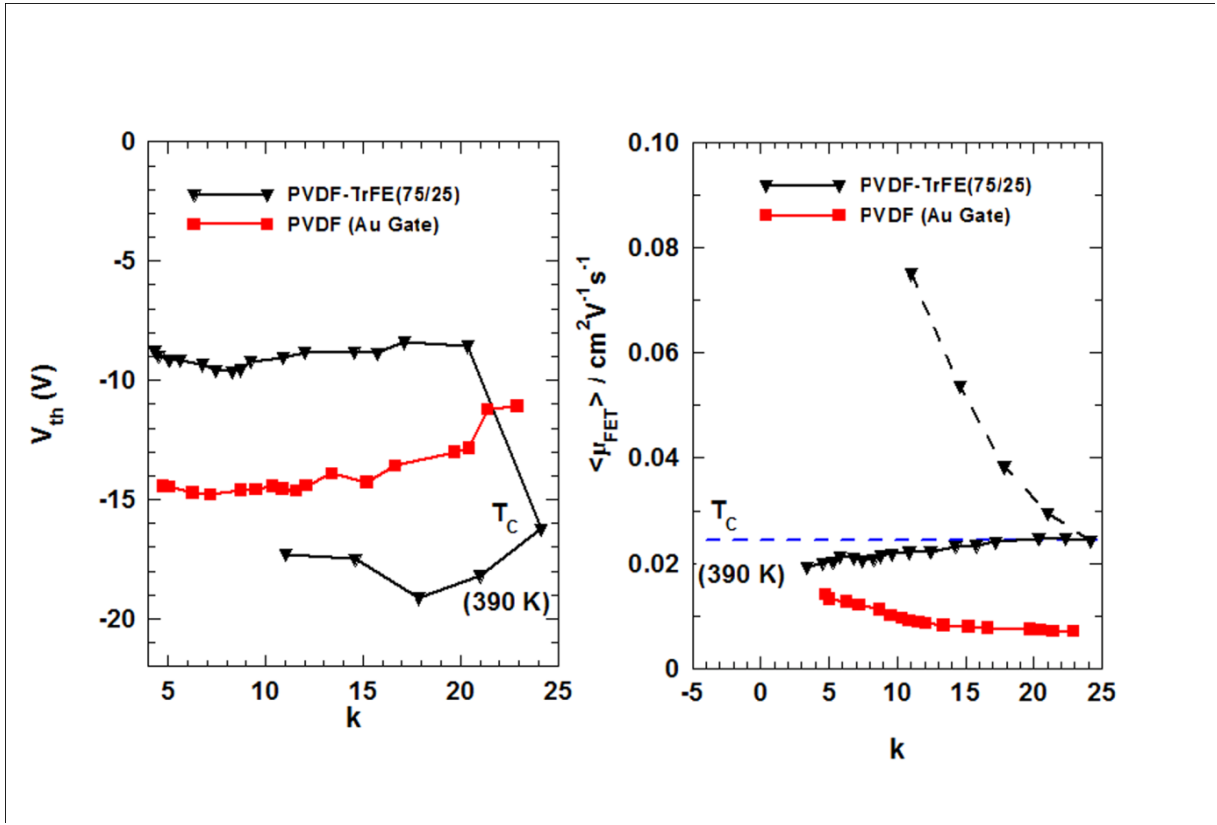

**Figure 5.** (a) Variation of $V_{th}$ with $k$ for FE-FET (PVDF/TrFE and β-PVDF as the dielectric); (b) variation of $<\mu_{FET}>$ with $k$ for the same device.

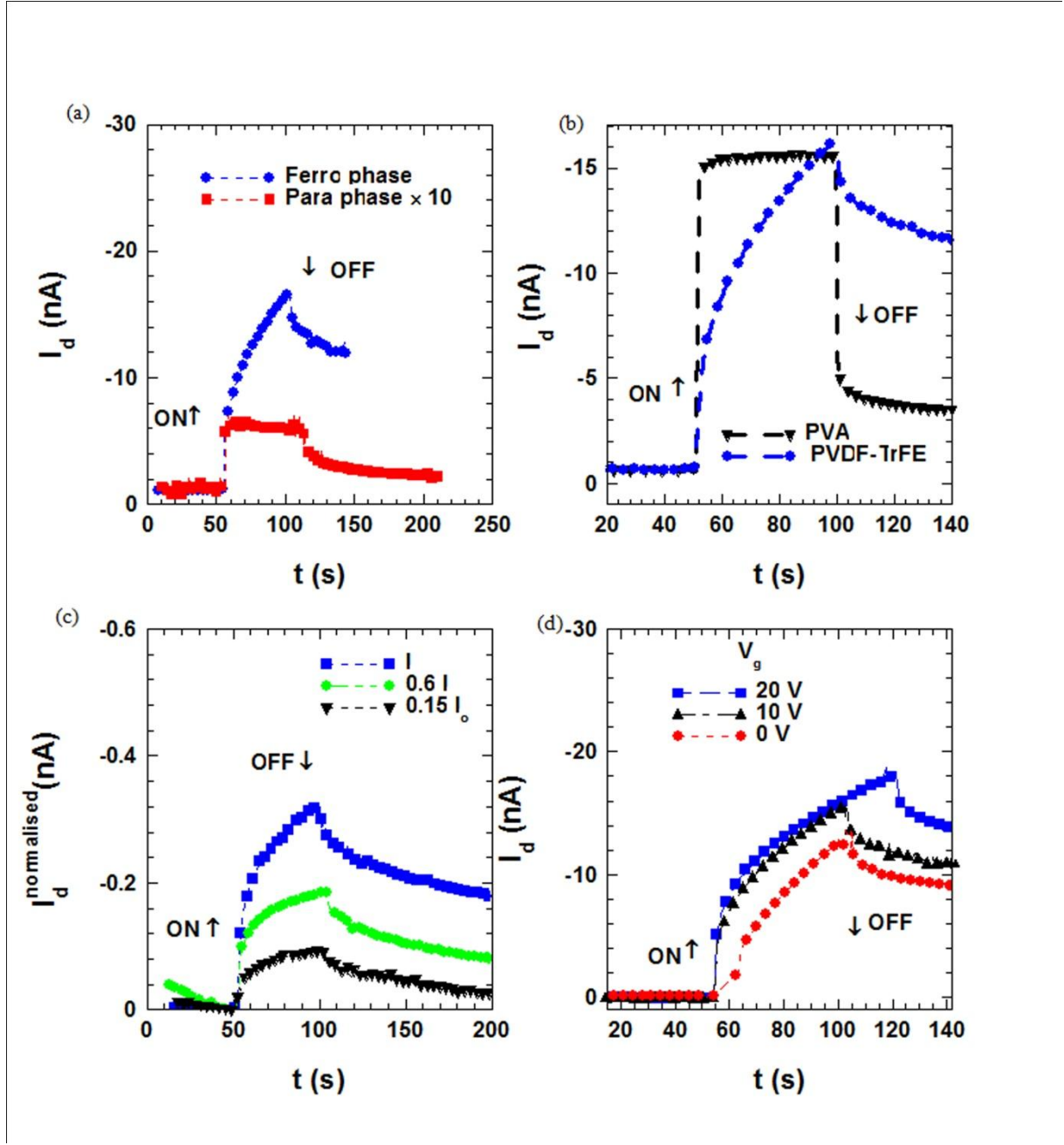

**Figure 6.** Photoresponse of the transistor with different dielectrics and P3HT as the semiconducting layer operating in the depletion mode ($V_d$ = -60 V, $V_g$ = 20 V) using a light source of intensity $I_o$ = 2 mWcm$^{-2}$ and $\lambda$ = 532 nm. (a) Difference of photo-response in the FE and PE phase. (b) Photo-response difference between P3HT-FETs made with high $k$ dielectric (PVA) and FE-dielectric PVDF-TrFE at 180 K. (c) Photo-response with variation of intensity of light showing different degree of retention. (d) Change in the photo-response with gate voltage showing the effect of tunable $\varepsilon_{dep}$ on $I_{ph}$ in FE-FET.